\documentclass{article}
\usepackage{spconf,amsmath,amssymb,graphicx,hyperref,url}
\usepackage{booktabs}
\usepackage{multirow}
\usepackage{xcolor}

\let\origunderscore\_
\newcommand{\metricus}{\origunderscore\allowbreak}
\DeclareRobustCommand{\metric}[1]{\texttt{\let\_\metricus #1}}

\title{Broadening Uncertainty Estimation for Audio Question Answering Across Methods, Formats, and Inputs}
\name{
Aaron Isidore Grace$^{1}$
\qquad
Weiran Wang$^{2}$
}

\address{
$^{1}$David R. Cheriton School of Computer Science, University of Waterloo, Waterloo, ON, Canada\\
$^{2}$Department of Computer Science, University of Iowa, Iowa City, IA, USA\\
aaron.grace@uwaterloo.ca \qquad weiran-wang@uiowa.edu
}

\begin{document}
\ninept
\maketitle

\begin{abstract}
Audio-language models can produce confident answers unsupported by the audio, motivating uncertainty estimates that identify unreliable responses. We compare probability-based, sampling-based, self-verification, evidential, and contrastive measures across four open-weight models and five audio QA benchmarks. In multiple-choice evaluation, first-token measures are strongest overall, with top-1 probability achieving a mean AUROC of .740, compared with .708 for ten-sample discrete semantic entropy, while requiring no additional model calls. Across four benchmarks, shifting from multiple-choice to open-ended evaluation lowers mean accuracy from 57.6\% to 36.6\%, yet uncertainty remains predictive of errors: semantic entropy, maximum token entropy, and semantic agreement achieve mean AUROCs of .697, .694, and .693, respectively. To test whether uncertainty reflects the evidence available to answer the question, we perform input ablations that remove either the audio or the question. Across top-1 confidence, entropy, and sampling-based measures, removing audio reduces error-detection AUROC by .101 on average, compared with .010 when removing the question. Together, these results establish efficient uncertainty baselines and show that uncertainty in audio-language models depends substantially more on available audio evidence than on question text.
\end{abstract}


\begin{table*}[t]
\centering
\fontsize{9}{10.5}\selectfont
\dimen0=\aboverulesep \dimen2=\belowrulesep
\setlength{\aboverulesep}{0.25ex}\setlength{\belowrulesep}{0.25ex}
\begin{tabular*}{\textwidth}{@{\extracolsep{\fill}}llrrrrrrrrrr@{}}
\toprule
& & & \multicolumn{3}{c}{\textbf{Single-pass}} & \multicolumn{1}{c}{\textbf{Sampling}} & \multicolumn{3}{c}{\textbf{Contrastive}} & \multicolumn{2}{c}{\textbf{Self-verification}} \\
\cmidrule(lr){4-6}\cmidrule(lr){7-7}\cmidrule(lr){8-10}\cmidrule(lr){11-12}
Benchmark & Model & Acc. & \emph{TOP1} & $H_{\mathrm{all}}$ & $\kappa$ & \emph{DSE} & \emph{UE} & \emph{Sup} & \emph{L2} & \emph{P(IK)} & \emph{P(aud)} \\
\midrule
\multirow{4}{*}{AQUA-Bench}
& AF3         & 55.5 & .685 & .660 & .642 & .646 & .665 & .596 & .564 & .487 & \textbf{.829} \\
& Phi-4-MM    & 31.2 & .510 & .499 & .392 & .504 & .514 & \textbf{.536} & .470 & .338 & .270 \\
& Qwen2-A     & 39.0 & .730 & .734 & .693 & .690 & .734 & \textbf{.746} & .685 & .659 & .575 \\
& Qwen2.5-O   & 73.9 & .729 & .692 & .693 & .703 & .702 & .507 & .413 & .458 & \textbf{.910} \\
\midrule
\multirow{4}{*}{MMAU}
& AF3         & 74.6 & \textbf{.860} & .851 & .842 & .823 & .856 & .651 & .492 & .604 & .623 \\
& Phi-4-MM    & 62.7 & \textbf{.788} & .784 & .567 & .752 & .771 & .551 & .434 & .572 & .593 \\
& Qwen2-A     & 65.8 & .804 & \textbf{.805} & .793 & .755 & \textbf{.805} & .699 & .578 & .568 & .653 \\
& Qwen2.5-O   & 75.3 & \textbf{.853} & .846 & .839 & .815 & .848 & .693 & .570 & .582 & .588 \\
\midrule
\multirow{4}{*}{MMAR}
& AF3         & 59.3 & \textbf{.695} & .693 & .671 & .675 & .681 & .647 & .579 & .526 & .547 \\
& Phi-4-MM    & 47.7 & .683 & \textbf{.694} & .579 & .680 & .661 & .601 & .577 & .467 & .548 \\
& Qwen2-A     & 48.2 & .681 & \textbf{.685} & .678 & .666 & .679 & .631 & .534 & .470 & .552 \\
& Qwen2.5-O   & 60.2 & \textbf{.742} & .726 & .704 & .697 & .727 & .712 & .647 & .510 & .555 \\
\midrule
\multirow{4}{*}{MMSU}
& AF3         & 58.4 & \textbf{.794} & .788 & .773 & .758 & \textbf{.794} & .774 & .667 & .496 & .537 \\
& Phi-4-MM    & 54.3 & \textbf{.770} & \textbf{.770} & .567 & .741 & .756 & .688 & .597 & .650 & .631 \\
& Qwen2-A     & 54.1 & .758 & \textbf{.760} & .757 & .721 & .748 & .679 & .569 & .595 & .603 \\
& Qwen2.5-O   & 61.7 & \textbf{.821} & \textbf{.821} & .816 & .794 & .810 & .737 & .648 & .659 & .525 \\
\midrule
\multirow{4}{*}{SAKURA}
& AF3         & 65.4 & .771 & .753 & .672 & .710 & .765 & \textbf{.863} & .774 & .574 & .555 \\
& Phi-4-MM    & 39.5 & .615 & .627 & .559 & .612 & .584 & \textbf{.633} & .578 & .483 & .494 \\
& Qwen2-A     & 58.0 & .704 & .705 & .701 & .672 & .735 & \textbf{.769} & .677 & .564 & .550 \\
& Qwen2.5-O   & 69.8 & .806 & .803 & .772 & .754 & .819 & \textbf{.832} & .775 & .687 & .737 \\
\bottomrule
\end{tabular*}
\setlength{\aboverulesep}{\dimen0}\setlength{\belowrulesep}{\dimen2}
\caption{Multiple-choice accuracy (\%) and uncertainty AUROC across
models and benchmarks. \textbf{Bold} indicates the
best AUROC in each row.}
\label{tab-main}
\end{table*}
\section{Introduction}
\label{sec-introduction}
A confident-sounding answer from an audio-language model is not necessarily grounded in the audio. Models can invent sounds, speakers, or events, or answer questions the recording cannot resolve \cite{kuan2024understanding,zhao2026halluaudio}. Because such hallucinations can sound as fluent as correct answers, identifying unreliable responses before they are acted on is essential. Uncertainty estimation offers one solution.\footnote{Code:\urlstyle{same}\url{https://github.com/aarongrace/uncertainty2026}}

Uncertainty estimation in language and vision-language models draws on
model probabilities and entropy \cite{malinin2020uncertainty},
variation across sampled answers \cite{farquhar2024semantic},
and explicit self-evaluation \cite{kadavath2022language}.
Recent approaches also measure evidential conflict \cite{huang2025evidential}
or changes in predictions when an input modality is removed or altered
\cite{park2026vauq,zhang2024vluncertainty}.
And in ASR, word-level confidence estimation is used to detect
transcription errors \cite{oneata2021evaluation}.

While uncertainty estimation is widely studied in text and vision models, its application to audio-language models is underexplored. Kuan et
al.~\cite{kuan2026walking} provide the first systematic comparison of
uncertainty measures with LALMs, while recent contrastive-decoding methods
use uncertainty to determine when intervention is needed
\cite{hsu2025audioaware,jung2025avcd,li2026temporal}. However, many recent LLM and VLM uncertainty methods remain untested in audio-language models.

In addition, most audio-language benchmarks use multiple-choice questions, yet this format has come under increasing scrutiny \cite{li2024canmcq,lin2025hearing}. Chandak et al.~\cite{chandak2025answermatching} show that answer choices provide strong cues even when the question is withheld. Their answer-matching method scores free-form responses against reference answers and agrees well with human judgments. ORCA \cite{sedlacek2025orca} evaluates open-ended audio responses using reference answers and textual audio grounding. However, whether uncertainty estimates remain reliable in open-ended settings is still unclear.

A related question is whether uncertainty reflects reliance on audio. Foo et al.~\cite{foo2026glitters} find that models retain 60–72\% of their performance without audio, suggesting substantial linguistic and answer-choice priors. However, how uncertainty changes when either the audio or the question is removed has not been measured.

We address these gaps by evaluating uncertainty across multiple-choice and
open-ended audio question answering and under controlled removal of input
information. Our contributions are:

\begin{itemize}
\item We conduct a broad comparison of uncertainty measures for audio-language models, spanning probability-based, sampling-based, self-verification, evidential, and contrastive signals. Across multiple-choice audio QA, simple first-token probability measures outperform more computationally expensive sampling and self-verification methods on average while requiring no additional model calls, establishing them as a strong and efficient baseline for uncertainty estimation.

\item We convert four multiple-choice benchmarks to open-ended evaluation by removing the answer choices and scoring free-form responses with an answer-matching model. Although accuracy changes substantially, uncertainty remains predictive of correctness, with sequence-level measures showing greater robustness across answer formats.

\item We separately remove the audio or the question while retaining the answer choices. Audio removal causes a much larger decline in uncertainty discrimination and greater changes in the ranking of uncertainty methods than question removal.

\end{itemize}

\section{Multiple-Choice Experimental Setup}
\label{sec-multiple-choice-setup}

\begin{table*}[t]
\centering
\begingroup
\fontsize{9}{10.5}\selectfont
\setlength{\tabcolsep}{2.5pt}
\setlength{\aboverulesep}{0.25ex}
\setlength{\belowrulesep}{0.25ex}
\begin{tabular*}{\textwidth}{@{\extracolsep{\fill}}ll*{12}{r}@{}}
\toprule
& & & \multicolumn{3}{c}{\textbf{Token uncertainty}} & \multicolumn{2}{c}{\textbf{Seq.\ likelihood}} & \multicolumn{2}{c}{\textbf{Sampling}} & \multicolumn{2}{c}{\textbf{Contrastive}} & \multicolumn{2}{c}{\textbf{Self-verification}} \\
\cmidrule(lr){4-6}\cmidrule(lr){7-8}\cmidrule(lr){9-10}\cmidrule(lr){11-12}\cmidrule(lr){13-14}
Benchmark & Model & Match & $\mathit{TOP1}_{\mathit{MIN}}$ & $H_{\max}$ & $H_{\mathrm{first}}$ & \emph{NLL} & \emph{PPL} & \emph{SemEnt} & \emph{SemAgr} & \emph{UE} & \emph{CWO} & \emph{P(IK)} & \emph{P(aud)} \\
\midrule
\multirow{4}{*}{AQUA}
& AF3         & 78.8 & .750 & .743 & .703 & .754 & .730 & .730 & \textbf{.758} & .666 & .376 & .489 & .378 \\
& Phi-4-MM    & 45.7 & .763 & .733 & .728 & \textbf{.782} & .778 & .759 & .759 & .718 & .242 & .281 & .281 \\
& Qwen2-A     & 32.5 & .715 & .776 & .762 & .732 & .716 & .771 & .755 & \textbf{.785} & .486 & .747 & .639 \\
& Qwen2.5-O   & 80.5 & .657 & .643 & .636 & .677 & .671 & .648 & .650 & .551 & .351 & .416 & \textbf{.718} \\
\midrule
\multirow{4}{*}{MMAR}
& AF3         & 31.5 & .692 & .687 & .586 & .678 & .621 & .697 & \textbf{.720} & .597 & .522 & .510 & .520 \\
& Phi-4-MM    & 25.6 & \textbf{.738} & .725 & .698 & .698 & .660 & .731 & .724 & .673 & .440 & .470 & .445 \\
& Qwen2-A     & 25.2 & .654 & .671 & .638 & .645 & .624 & \textbf{.675} & .668 & .641 & .464 & .502 & .521 \\
& Qwen2.5-O   & 30.6 & .712 & .711 & .695 & .698 & .677 & \textbf{.719} & .696 & .700 & .506 & .530 & .632 \\
\midrule
\multirow{4}{*}{MMAU}
& AF3         & 35.1 & .655 & .659 & .697 & .571 & .676 & .668 & .662 & \textbf{.707} & .444 & .613 & .583 \\
& Phi-4-MM    & 25.6 & .695 & .724 & .735 & .624 & .692 & .688 & .671 & \textbf{.739} & .453 & .449 & .581 \\
& Qwen2-A     & 26.1 & \textbf{.641} & .640 & .635 & .600 & .639 & .635 & .633 & .629 & .492 & .625 & .586 \\
& Qwen2.5-O   & 30.3 & .688 & \textbf{.712} & .708 & .636 & .684 & .705 & .663 & .709 & .422 & .612 & .618 \\
\midrule
\multirow{4}{*}{MMSU}
& AF3         & 32.4 & .699 & .669 & .636 & .650 & .648 & .685 & \textbf{.700} & .617 & .585 & .469 & .527 \\
& Phi-4-MM    & 27.1 & .633 & .643 & .634 & .586 & \textbf{.679} & .665 & .668 & .619 & .468 & .600 & .563 \\
& Qwen2-A     & 24.7 & .648 & .667 & .659 & .627 & .678 & \textbf{.684} & .681 & .647 & .517 & .614 & .533 \\
& Qwen2.5-O   & 34.0 & \textbf{.709} & .707 & .707 & .620 & .708 & .684 & .673 & .705 & .526 & .674 & .656 \\
\bottomrule
\end{tabular*}
\endgroup
\caption{Open-ended evaluation: judge-match accuracy (\%) and uncertainty AUROC. \textbf{Bold} indicates the best AUROC in each row.}
\label{tab-oe}
\end{table*}

We evaluate four LALMs: Audio Flamingo 3 (AF3) \cite{goel2025audioflamingo3}, Phi-4-multimodal-instruct (Phi-4-MM) \cite{abouelenin2025phi4}, Qwen2-Audio-7B-Instruct (Qwen2-A) \cite{chu2024qwen2}, and Qwen2.5-Omni-7B (Qwen2.5-O) \cite{xu2025qwen25omni}. We test five audio QA benchmarks. \textbf{AQUA-Bench} includes standard QA and abstention cases with insufficient evidence, missing correct answers, or incompatible answer sets \cite{kuan2026aqua}. \textbf{MMAU} spans speech, environmental sounds, and music; we use its 1{,}000-question test-mini split \cite{sakshi2025mmau}. \textbf{MMAR} targets multi-step audio reasoning \cite{ma2025mmar}, \textbf{MMSU} fine-grained linguistic and paralinguistic understanding \cite{wang2025mmsu}, and \textbf{SAKURA} matched single- and multi-hop reasoning \cite{yang2025sakura}. Models are prompted to output the option letter first.

Because models can exhibit option-position bias \cite{lin2025hearing,zheng2024selectionbias}, we place the correct answer once in each available position and average across the permutations. Particuarly, this matters for AQUA-Bench, which places abstention at option E. AF3 achieves only 19.7\% accuracy at E, versus 54.5\% across A--D, suggesting that previously reported low performance \cite{kuan2026walking,kuan2026aqua} may reflect position bias.

\section{Multiple-Choice Uncertainty}
\label{sec-mcq}

We use each uncertainty score to distinguish incorrect from correct answers, treating errors as the positive class. Performance is measured by area under the receiver operating characteristic curve (AUROC), where higher values indicate better separation and .5 corresponds to chance. Confidence-based metrics are inverted so that higher scores consistently indicate greater uncertainty.

\smallskip\noindent\textbf{First token distribution.}
Let $p$ be the first-token distribution over vocabulary $V$. We treat top-1 token probability as confidence and normalized full-vocabulary entropy as uncertainty:
\begin{equation}
\mathit{TOP1}=\max_{v\in V}p_v,\qquad
H_{\mathrm{all}}=\frac{-\sum_{v\in V}p_v\log p_v}{\log|V|}.
\label{eq:option-posterior}
\end{equation}
Despite their simplicity, first-token scores are the strongest overall methods in Table~\ref{tab-main}. Much of the useful uncertainty signal is therefore already available at the point of decision, without additional sampling or verification. Because these scores are available before further generation, they are more suitable for latency-sensitive audio applications where uncertainty must be assessed in real time.

\smallskip\noindent\textbf{Evidential conflict.}
Following Huang et al.~\cite{huang2025evidential}, we center output weights across answer options and write each resulting logit as \(w_i^+-w_i^-\). Here, \(w_i^+\) and \(w_i^-\) sum the positive and negative contributions from final hidden-state features by magnitude. Retaining both sums exposes evidence that would otherwise cancel in the final logit. Let \(\Omega=\{o_1,\ldots,o_M\}\) denote the \(M\) answer options. A Dempster--Shafer mass function \(m:2^\Omega\to[0,1]\) assigns mass \(m(A)\) to each subset \(A\subseteq\Omega\), representing support that the correct answer lies in \(A\), with \(m(\varnothing)=0\) and \(\sum_{A\subseteq\Omega}m(A)=1\). Positive evidence assigns mass \(1-e^{-w_i^+}\) to \(\{o_i\}\), while negative evidence assigns \(1-e^{-w_i^-}\) to \(\Omega\setminus\{o_i\}\); remaining mass is assigned to \(\Omega\).

Combining these masses across options separately for each sign using Dempster's rule gives \(m^+\) and \(m^-\). Their conflict is
\begin{equation}
\kappa=\sum_{\substack{B,C\subseteq\Omega\\B\cap C=\varnothing}}m^+(B)m^-(C).
\end{equation}
Thus, \(\kappa\) measures how much positive and negative evidence supports incompatible answer sets. A high \(\kappa\) indicates strong contradictory evidence within the model rather than merely weak evidence overall. Its competitive AUROC, distinct mechanism, and lack of additional model calls motivate further study, although large evidence magnitudes can cause saturation.

\smallskip\noindent\textbf{Sampling.}
Discrete semantic entropy (DSE) measures disagreement among $K=10$ sampled
answers \cite{kuan2026walking}.
\begin{equation}
\hat p_j=\frac{1}{K}\sum_{k=1}^{K}\mathbf{1}[y^{(k)}=j],
\qquad \mathit{DSE}=H(\hat p).
\end{equation}
Despite requiring multiple generations, \emph{DSE} achieves a mean AUROC of .708, below $H_{\mathrm{all}}$ (.735) and \emph{TOP1} (.740) in Table~\ref{tab-main}.

\smallskip\noindent\textbf{Audio--silence contrast.}
Alongside the answering pass, we run one further single-token pass with the
audio replaced by silence and read its distribution the same way.
Renormalizing over the options gives audio/silence posteriors $c^a,c^0$
(full-vocabulary $p^a,p^0$); $y$ is the option chosen under real audio. This
asks not how confident the model is, but how much of that confidence the
audio is responsible for.

\emph{Weighted support (Sup).} Whether the audio backs the model's specific
answer, and whether the model believes it: the chosen option's audio-branch
probability, scaled by its gain over silence.

\emph{Unexplained entropy (UE).} Exactly VAUQ's $s_{\mathrm{VAUQ}}$
\cite{park2026vauq}, read over the option posterior and built on a
whole-clip silence ablation rather than VAUQ's core-region masking:
predictive uncertainty discounted by however much of it the audio explains.

\emph{L2} is the full-distribution shift between audio and silence.

\begin{equation}
\begin{aligned}
\mathit{Sup} &= c^a_y(c^a_y-c^0_y), \\
\mathit{UE} &= H(c^a)-\alpha[H(c^0)-H(c^a)],\quad \alpha=.5, \\
\mathit{L2} &= \|p^a-p^0\|_2/\sqrt2.
\end{aligned}
\end{equation}

Among the audio--silence metrics, \emph{UE} performs most consistently
overall, while \emph{Sup} performs especially well on SAKURA. \emph{L2} is
generally weaker: it keeps the size of the distributional shift but not its
direction, discarding exactly the graded, signed information \emph{UE} and
\emph{Sup} are built to preserve.

\smallskip\noindent\textbf{Self-verification.}

A second pass evaluates the answer \(\hat y\) under a verification prompt \(\pi\), using the probability of affirmation as confidence \cite{kadavath2022language,kuan2026walking}:
\[
\mathit{SV}=p_\theta(z^+\mid a,q,\hat y,\pi).
\]

We test several prompts and report two in Table~\ref{tab-main}: \emph{P(IK)}, following the published baseline \cite{kuan2026walking}, and \emph{P(aud)}, which asks whether the audio supports the answer. Thus, \emph{P(aud)} reflects the model's stated audio support rather than measured audio dependence. Overall, self-verification is weaker and less stable than first-token confidence, though \emph{P(aud)} performs well in some settings.

We also evaluated Qwen2.5-Omni-3B, which showed similar uncertainty trends but lower mean accuracy than the 7B model (65.2\% vs. 68.2\%) and AUROC lower by .011--.022 across metrics. We therefore report only the 7B results for brevity.

\begin{table*}[t]
\centering
\begingroup
\fontsize{9}{10.5}\selectfont
\setlength{\tabcolsep}{2.5pt}
\setlength{\aboverulesep}{0.25ex}
\setlength{\belowrulesep}{0.25ex}
\begin{tabular*}{\textwidth}{@{\extracolsep{\fill}}l*{12}{r}@{}}
\toprule
& \multicolumn{4}{c}{\textbf{Canonical}} & \multicolumn{4}{c}{\textbf{No question}} & \multicolumn{4}{c}{\textbf{No audio}} \\
\cmidrule(lr){2-5}\cmidrule(lr){6-9}\cmidrule(lr){10-13}
& \multicolumn{2}{c}{$H_{\mathrm{all}}$} & \multicolumn{2}{c}{\emph{DSE}}
& \multicolumn{2}{c}{$H_{\mathrm{all}}$} & \multicolumn{2}{c}{\emph{DSE}}
& \multicolumn{2}{c}{$H_{\mathrm{all}}$} & \multicolumn{2}{c}{\emph{DSE}} \\
Dataset & Mean score & AUROC & Mean score & AUROC & $\Delta$ & AUROC & $\Delta$ & AUROC & $\Delta$ & AUROC & $\Delta$ & AUROC \\
\midrule
AQUA-Bench & .068 & \textbf{.646} & .610 & .636 & $+.004$ & \textbf{.667} & $+.036$ & .649 & $+.018$ & \textbf{.630} & $+.148$ & .613 \\
MMAU & .049 & \textbf{.821} & .460 & .786 & $+.010$ & \textbf{.786} & $+.096$ & .756 & $+.016$ & \textbf{.728} & $+.139$ & .699 \\
MMAR & .063 & \textbf{.700} & .588 & .679 & $+.003$ & \textbf{.707} & $+.025$ & .684 & $+.012$ & \textbf{.606} & $+.100$ & .575 \\
MMSU & .057 & \textbf{.785} & .537 & .753 & $+.004$ & \textbf{.756} & $+.029$ & .729 & $+.015$ & \textbf{.659} & $+.120$ & .638 \\
SAKURA & .055 & \textbf{.722} & .493 & .687 & $+.008$ & \textbf{.709} & $+.052$ & .674 & $+.021$ & \textbf{.549} & $+.147$ & .545 \\
\bottomrule
\end{tabular*}
\endgroup
\caption{Multiple-choice input ablation, averaged over four models. \emph{Mean score}
is the mean uncertainty value under full input, $\Delta$ its mean change under
ablation (ablated minus canonical). \textbf{Bold} marks the best AUROC in each
condition.}
\label{tab-ablation}
\end{table*}

\section{Open-Ended Uncertainty}
\label{sec-oe}

We convert the multiple-choice benchmarks to open-ended evaluation by
removing the answer choices. SAKURA is excluded because many of its
questions depend on the answer choices. For MMAU and MMAR, we
exclude questions that cannot be answered without the option menu,
dropping 78 of MMAU's 1{,}000 questions and 23 of MMAR's 995.
MMSU is filtered to exclude 11 menu-dependent task families out of 47,
removing 1{,}181 of its 5{,}000 questions. For AQUA-Bench, we retain the
original and mismatched audio--question subsets, with an abstention hint 
in the prompt for the latter. A Qwen3-4B judge scores responses against
the reference using Chandak et al.'s~\cite{chandak2025answermatching}
answer-matching protocol, marking a response correct when it conveys at
least as much information as the reference. We evaluate open-ended
uncertainty measures that generally correspond to the multiple-choice
metrics, allowing a practical comparison across answer formats. Results
are reported in Table~\ref{tab-oe}.

Removing answer choices generally lowers accuracy: mean
accuracy falls from 69.6\% to 29.3\% on MMAU, from 57.1\% to 29.5\% on
MMSU, and from 53.9\% to 28.2\% on MMAR. However, accuracy rises from
49.9\% to 59.4\% on AQUA-Bench. This suggests that models can more easily
express abstention in free-form responses without the distraction of
answer labels.

\smallskip\noindent\textbf{Token confidence and entropy.}
For the vocabulary distribution $p_t$ at generation step $t$, we compare
first-token entropy with the most uncertain token in the response:
\begin{equation}
\begin{aligned}
\mathit{TOP1}_{\mathit{MIN}}&=\min_t\max_{v\in V}p_t(v), &
H_{\max}&=\max_t H(p_t),\\
H_{\mathrm{first}}&=H(p_1).
\end{aligned}
\end{equation}
$H_{\mathrm{first}}$ uses entropy at the first token. $H_{\max}$ takes the maximum entropy and $TOP1_{\min}$ the minimum top-1 probability across the response. $H_{\mathrm{first}}$ performs worse overall, suggesting that later tokens provide additional uncertainty information.

\smallskip\noindent\textbf{Sequence likelihood.}
Negative log-likelihood (\emph{NLL}) and perplexity (\emph{PPL}) measure uncertainty from the probabilities assigned to the generated tokens. For an answer $y_{1:T}$,
\begin{equation}
NLL=-\sum_{t=1}^{T}\log p_t(y_t),\qquad
\mathit{PPL}=\exp(NLL/T).
\end{equation}
\emph{NLL} sums token surprise across the entire response and therefore grows with response length, whereas \emph{PPL} normalizes by length and measures average token surprise. Both perform competitively in Table~\ref{tab-oe}, with \emph{NLL} particularly strong on AQUA-Bench and \emph{PPL} on MMSU; neither dominates consistently overall.

\smallskip\noindent\textbf{Contrastive.}
We compare generation with audio, denoted by superscript ${}^{a}$, to generation
with the audio removed, denoted by superscript ${}^{0}$. Here, $p_1^{a}$ and
$p_1^{0}$ are the first-token distributions under the two conditions, while
$y^{a}$ and $y^{0}$ are the corresponding responses. Let $W(y)$ denote the
set of content words in response $y$. Unexplained entropy (\emph{UE}) and content-word overlap
(\emph{CWO}) are:
\begin{equation}
\begin{gathered}
\mathit{UE}
=H(p_1^{a})-\alpha\!\left[H(p_1^{0})-H(p_1^{a})\right],
\qquad \alpha=.5,\\[2pt]
\mathit{CWO}
=\frac{|W(y^{a})\cap W(y^{0})|}
{|W(y^{a})\cup W(y^{0})|}.
\end{gathered}
\end{equation}

\smallskip\noindent\textbf{Sampling.}
Semantic uncertainty from free-form responses accounts for different
wordings of the same answer \cite{kuhn2023semantic,farquhar2024semantic}.
We draw $K=10$ stochastic responses $y^{(1)},\dots,y^{(K)}$ and let $\hat y$
denote the greedily decoded response, obtained by selecting the
highest-probability token at each step. Using
DeBERTa-large-MNLI conditioned on the question, we treat two responses as
equivalent, $y^{(i)}\equiv y^{(j)}$, when they entail one another in both
directions. The sampled responses are grouped into semantic clusters
with labels $c_k\in\{1,\dots,C\}$.

\begin{equation}
\begin{gathered}
\hat p_j =
\frac{1}{K}\sum_{k=1}^{K}\mathbf{1}[c_k=j],
\qquad
\mathit{SemEnt} =
-\sum_{j=1}^{C}\hat p_j\log\hat p_j,\\[2pt]
\mathit{SemAgr} =
\frac{1}{K}\sum_{k=1}^{K}
\mathbf{1}\!\left[y^{(k)}\equiv\hat y\right].
\end{gathered}
\end{equation}

\emph{SemEnt} measures how dispersed the samples are across semantic
clusters, while \emph{SemAgr} measures how often sampled responses agree
with the greedy response. These measures are competitive overall:
\emph{SemEnt} or \emph{SemAgr} achieves the best AUROC in 6 of 16
model--benchmark pairs, including three of four MMAR rows and two of four
MMSU rows. Semantic grouping also avoids treating paraphrases as distinct
answers, making it better suited to free-form sampling than exact-string
\emph{DSE}.

\smallskip\noindent\textbf{Self-verification.}
Both self-verification prompts remain relatively weak in the open-ended setting. Mean \emph{P(aud)} AUROC decreases from .596 to .549, while \emph{P(IK)} remains nearly unchanged. Together with their modest multiple-choice performance, this suggests that the models have limited ability to reliably assess the correctness or audio support of their own answers through textual queries.

\section{Uncertainty with Missing Inputs}
\label{sec-ablations}

The question, audio, and answer choices can each provide clues to the answer. We remove either the audio or the question while retaining the remaining inputs to examine how uncertainty depends on these sources of evidence and whether it still identifies errors. Table~\ref{tab-ablation} averages
results across four models per dataset; overall means weight the 20
model--benchmark pairs equally. Higher mean $H_{\mathrm{all}}$ and \emph{DSE}
scores indicate greater model uncertainty.

\smallskip\noindent\textbf{Removing the question.}
Mean accuracy falls from 57.7\% to 54.2\%, while \emph{TOP1} and
$H_{\mathrm{all}}$ each lose only .010 AUROC. The scores barely move either:
$H_{\mathrm{all}}$ rises .006 and \emph{DSE} .048.
One possible explanation is that descriptive answer choices can make
the question unnecessary. For example, given options such as ``a dog
barking'' and ``a car horn,'' a model can select the sound heard in the
recording without seeing the question. With these cues available, removing
the question has little effect on uncertainty's ability to identify errors.

\smallskip\noindent\textbf{Removing audio.}
Removing audio has a larger effect than removing the question. Mean
accuracy falls by 17.8 versus 3.5 percentage points, and \emph{TOP1}
AUROC declines by .108 versus .010. Full-vocabulary entropy shows a
similar decline in error discrimination. The scores also respond more
strongly to missing audio. Mean $H_{\mathrm{all}}$ rises by .016 without
audio versus .006 without the question, with \emph{DSE} following the
same pattern. Thus, audio removal increases uncertainty while making
errors harder to identify. The smaller \emph{TOP1} decline on AQUA-Bench
may reflect questions that models can recognize as unanswerable from
text alone.

First-token confidence and entropy retain useful error discrimination
without audio, outperforming ten-sample \emph{DSE} on average.
Evidential conflict also becomes more competitive, narrowing its mean
AUROC gap with \emph{TOP1} from .054 to .006, although its absolute AUROC
declines for three of four models. Self-verification is more affected,
with \emph{P(aud)} approaching chance at .516 AUROC.

\section{Discussions}
\label{sec-discussion}
Existing approaches use uncertainty to gate chain-of-thought (CoT)
reasoning~\cite{kuan2026walking} or guide contrastive
decoding~\cite{li2026temporal}. We instead tried giving models their own
uncertainty scores to help them revise their answers. Each prompt included
the original question, the previous answer, and one score, presented alone
or with a calibrated summary or thirty worked examples. This approach did not improve accuracy. Numerical feedback reduced
accuracy by 10.4--12.3 percentage points, and asking models to reason
before revising made performance worse, with losses reaching 23.6 points.

These results suggest that the tested models do not reliably use their own uncertainty for self-correction. Future work could combine multiple uncertainty signals with a learned probe to produce a calibrated estimate of correctness. Bayesian decision theory could then turn this estimate into an explicit choice to retain, revise, or abstain, offering a principled way to balance answer quality against the costs of errors and additional inference.

{\small
\bibliographystyle{IEEEbib}
\bibliography{refs}

@inproceedings{malinin2020uncertainty,
  title     = {Uncertainty Estimation in Autoregressive Structured Prediction},
  author    = {Malinin, Andrey and Gales, Mark},
  booktitle = {ICLR},
  year      = {2021}
}

@inproceedings{kuhn2023semantic,
  title     = {Semantic Uncertainty: Linguistic Invariances for Uncertainty Estimation in Natural Language Generation},
  author    = {Kuhn, Lorenz and Gal, Yarin and Farquhar, Sebastian},
  booktitle = {ICLR},
  year      = {2023}
}

@article{farquhar2024semantic,
  title   = {Detecting Hallucinations in Large Language Models Using Semantic Entropy},
  author  = {Farquhar, Sebastian and Kossen, Jannik and Kuhn, Lorenz and Gal, Yarin},
  journal = {Nature},
  volume  = {630},
  number  = {8017},
  pages   = {625--630},
  year    = {2024}
}

@article{kadavath2022language,
  title   = {Language Models (Mostly) Know What They Know},
  author  = {Kadavath, Saurav and Conerly, Tom and Askell, Amanda and Henighan, Tom and Drain, Dawn and Perez, Ethan and Schiefer, Nicholas and Hatfield-Dodds, Zac and DasSarma, Nova and Tran-Johnson, Eli and others},
  journal = {arXiv preprint arXiv:2207.05221},
  year    = {2022}
}

@article{oneata2021evaluation,
  title   = {An Evaluation of Word-Level Confidence Estimation for End-to-End Automatic Speech Recognition},
  author  = {Oneata, Dan and Caranica, Alexandru and Stan, Adriana and Cucu, Horia},
  journal = {arXiv preprint arXiv:2101.05525},
  year    = {2021}
}

@article{chu2024qwen2,
  title   = {Qwen2-Audio Technical Report},
  author  = {Chu, Yunfei and Xu, Jin and Yang, Qian and Wei, Haojie and Wei, Xipin and Guo, Zhifang and Leng, Yichong and Lv, Yuanjun and He, Jinzheng and Lin, Junyang and others},
  journal = {arXiv preprint arXiv:2407.10759},
  year    = {2024}
}

@article{xu2025qwen25omni,
  title   = {Qwen2.5-Omni Technical Report},
  author  = {Xu, Jin and Guo, Zhifang and He, Jinzheng and Hu, Hangrui and He, Ting and Bai, Shuai and Chen, Keqin and Wang, Jialin and Fan, Yang and Dang, Kai and others},
  journal = {arXiv preprint arXiv:2503.20215},
  year    = {2025}
}

@article{goel2025audioflamingo3,
  title   = {Audio Flamingo 3: Advancing Audio Intelligence with Fully Open Large Audio Language Models},
  author  = {Goel, Arushi and Ghosh, Sreyan and Kim, Jaehyeon and Kumar, Sonal and Kong, Zhifeng and Lee, Sang-gil and Yang, Chao-Han Huck and others},
  journal = {arXiv preprint arXiv:2507.08128},
  year    = {2025}
}

@article{abouelenin2025phi4,
  title   = {Phi-4-Mini Technical Report: Compact yet Powerful Multimodal Language Models via Mixture-of-{LoRA}s},
  author  = {Abouelenin, Abdelrahman and Ashfaq, Atabak and Atkinson, Adam and Awadalla, Hany and Bach, Nguyen and Bao, Jianmin and Benhaim, Alon and Cai, Martin and Chaudhary, Vishrav and Chen, Congcong and others},
  journal = {arXiv preprint arXiv:2503.01743},
  year    = {2025}
}

@inproceedings{kuan2024understanding,
  title     = {Understanding Sounds, Missing the Questions: The Challenge of Object Hallucination in Large Audio-Language Models},
  author    = {Kuan, Chun-Yi and Huang, Wei-Ping and Lee, Hung-yi},
  booktitle = {Proceedings of Interspeech},
  year      = {2024}
}

@article{zhao2026halluaudio,
  title   = {{HalluAudio}: A Comprehensive Benchmark for Hallucination Detection in Large Audio-Language Models},
  author  = {Zhao, Feiyu and Chen, Yiming and Lu, Wenhuan and Zhang, Daipeng and Yue, Xianghu and Wei, Jianguo},
  journal = {arXiv preprint arXiv:2604.19300},
  year    = {2026}
}

@article{hsu2025audioaware,
  title   = {Reducing Object Hallucination in Large Audio-Language Models via Audio-Aware Decoding},
  author  = {Hsu, Tzu-wen and Lu, Ke-Han and Chiang, Cheng-Han and Lee, Hung-yi},
  journal = {arXiv preprint arXiv:2506.07233},
  year    = {2025}
}

@inproceedings{kuan2026aqua,
  title     = {{AQUA-Bench}: Beyond Finding Answers to Knowing When There Are None in Audio Question Answering},
  author    = {Kuan, Chun-Yi and Lee, Hung-yi},
  booktitle = {ICASSP},
  year      = {2026}
}

@article{foo2026glitters,
  title   = {All That Glitters Is Not Audio: Rethinking Text Priors and Audio Reliance in Audio-Language Evaluation},
  author  = {Foo, Leonardo Haw-Yang and Yang, Chih-Kai and Li, Chen-An and Lu, Ke-Han and Lee, Hung-yi},
  journal = {arXiv preprint arXiv:2604.24401},
  year    = {2026}
}

@article{zhang2024vluncertainty,
  title   = {{VL-Uncertainty}: Detecting Hallucination in Large Vision-Language Model via Uncertainty Estimation},
  author  = {Zhang, Ruiyang and Zhang, Hu and Zheng, Zhedong},
  journal = {arXiv preprint arXiv:2411.11919},
  year    = {2024}
}

@article{li2026temporal,
  title   = {Temporal Contrastive Decoding: A Training-Free Method for Large Audio-Language Models},
  author  = {Li, Yanda and Liu, Yuhan and Song, Zirui and Wei, Yunchao and Tak\'a\v{c}, Martin and Lahlou, Salem},
  journal = {arXiv preprint arXiv:2604.15383},
  year    = {2026}
}

@article{kuan2026walking,
  title   = {Walking Through Uncertainty: An Empirical Study of Uncertainty Estimation for Audio-Aware Large Language Models},
  author  = {Kuan, Chun-Yi and Huang, Wei-Ping and Lee, Hung-yi},
  journal = {arXiv preprint arXiv:2604.25591},
  year    = {2026}
}

@inproceedings{zheng2024selectionbias,
  title     = {Large Language Models Are Not Robust Multiple Choice Selectors},
  author    = {Zheng, Chujie and Zhou, Hao and Meng, Fandong and Zhou, Jie and Huang, Minlie},
  booktitle = {ICLR},
  year      = {2024}
}

@article{lin2025hearing,
  title   = {Hearing the Order: Investigating Position Bias in Large Audio-Language Models},
  author  = {Lin, Yu-Xiang and Li, Chen-An and Wei, Sheng-Lun and Chen, Po-Chun and Chen, Hsin-Hsi and Lee, Hung-yi},
  journal = {arXiv preprint arXiv:2510.00628},
  year    = {2025}
}

@inproceedings{li2024canmcq,
  title     = {Can Multiple-Choice Questions Really Be Useful in Detecting the Abilities of {LLM}s?},
  author    = {Li, Wangyue and Li, Liangzhi and Xiang, Tong and Liu, Xiao and Deng, Wei and Garcia, Noa},
  booktitle = {Proceedings of LREC-COLING},
  year      = {2024}
}

@article{chandak2025answermatching,
  title   = {Answer Matching Outperforms Multiple Choice for Language Model Evaluation},
  author  = {Chandak, Nikhil and Goel, Shashwat and Prabhu, Ameya and Hardt, Moritz and Geiping, Jonas},
  journal = {arXiv preprint arXiv:2507.02856},
  year    = {2025}
}

@inproceedings{sakshi2025mmau,
  title     = {{MMAU}: A Massive Multi-Task Audio Understanding and Reasoning Benchmark},
  author    = {Sakshi, S and Tyagi, Utkarsh and Kumar, Sonal and Seth, Ashish and Selvakumar, Ramaneswaran and Nieto, Oriol and Duraiswami, Ramani and Ghosh, Sreyan and Manocha, Dinesh},
  booktitle = {ICLR},
  year      = {2025}
}

@article{ma2025mmar,
  title   = {{MMAR}: A Challenging Benchmark for Deep Reasoning in Speech, Audio, Music, and Their Mix},
  author  = {Ma, Ziyang and Ma, Yinghao and Zhu, Yanqiao and Yang, Chen and Chao, Yi-Wen and Xu, Ruiyang and Chen, Wenxi and Chen, Yuanzhe and Chen, Zhuo and Cong, Jian and others},
  journal = {arXiv preprint arXiv:2505.13032},
  year    = {2025}
}

@article{wang2025mmsu,
  title   = {{MMSU}: A Massive Multi-task Spoken Language Understanding and Reasoning Benchmark},
  author  = {Wang, Dingdong and Li, Junan and Wu, Jincenzi and Yang, Dongchao and Chen, Xueyuan and Zhang, Tianhua and Meng, Helen},
  journal = {arXiv preprint arXiv:2506.04779},
  year    = {2025}
}

@article{yang2025sakura,
  title   = {{SAKURA}: On the Multi-hop Reasoning of Large Audio-Language Models Based on Speech and Audio Information},
  author  = {Yang, Chih-Kai and Ho, Neo and Piao, Yen-Ting and Lee, Hung-yi},
  journal = {arXiv preprint arXiv:2505.13237},
  year    = {2025}
}

@inproceedings{park2026vauq,
  title     = {{VAUQ}: Vision-Aware Uncertainty Quantification for {LVLM} Self-Evaluation},
  author    = {Park, Seongheon and Oh, Changdae and Choi, Hyeong Kyu and Du, Sean and Li, Sharon},
  booktitle = {Findings of ACL},
  year      = {2026}
}

@inproceedings{jung2025avcd,
  title     = {{AVCD}: Mitigating Hallucinations in Audio-Visual Large Language Models through Contrastive Decoding},
  author    = {Jung, Chaeyoung and Jang, Youngjoon and Chung, Joon Son},
  booktitle = {NeurIPS},
  year      = {2025}
}

@article{huang2025evidential,
  title   = {Visual Hallucination Detection in Large Vision-Language Models via Evidential Conflict},
  author  = {Huang, Tao and Liu, Zhekun and Wang, Rui and Zhang, Yang and Jing, Liping},
  journal = {arXiv preprint arXiv:2506.19513},
  year    = {2025}
}

@article{sedlacek2025orca,
  title   = {{ORCA}: Open-ended Response Correctness Assessment for Audio Question Answering},
  author  = {Sedl{\'a}{\v{c}}ek, {\v{S}}imon and Barahona, Sara and Yusuf, Bolaji and Herrera-Alarc{\'o}n, Laura and Kesiraju, Santosh and Bola{\~n}os, Cecilia and Lozano-Diez, Alicia and Udupa, Sathvik and L{\'o}pez, Fernando and others},
  journal = {arXiv preprint arXiv:2512.09066},
  year    = {2025}
}
}

\section{Compliance with Ethical Standards}
This study used publicly available data from AQUA-Bench, MMAU, MMAR, MMSU, and SAKURA, together with publicly released open-weight models. No new human- or animal-subject data were collected; therefore, ethical approval was not required. 
Weiran Wang is supported by a Google Gift Award. The authors declare no conflicts of interest.
\end{document}